# Polar vortex hidden in twisted bilayers of paraelectric SrTiO$_3$

Haozhi Sha[1,2,3,†], Yixuan Zhang[4,†], Yunpeng Ma[1,3], Wei Li[1,3], Wenfeng Yang[1,2,3], Jizhe Cui[1,2,3], Qian Li[1,3,*], Houbing Huang[4,*], Rong Yu[1,2,3,*]

**Affiliations:**

[1]School of Materials Science and Engineering, Tsinghua University, Beijing 100084, China.

[2]MOE Key Laboratory of Advanced Materials, Tsinghua University, Beijing 100084, China.

[3]State Key Laboratory of New Ceramics and Fine Processing, Tsinghua University, Beijing 100084, China.

[4]Advanced Research Institute of Multidisciplinary Science & School of Materials Science and Engineering, Beijing Institute of Technology, Beijing 100081, China.

*Corresponding author. Email: ryu@tsinghua.edu.cn, hbhuang@bit.edu.cn, qianli_mse@tsinghua.edu.cn

**Polar topologies, such as vortex and skyrmion, have attracted significant interest due to their unique physical properties and promising applications in high-density memory devices. Currently, most polar vortices are observed in heterostructures containing ferroelectric materials and constrained by substrates. In this study, we unravel arrays of polar vortices formed in twisted freestanding bilayers composed of SrTiO$_3$, a quantum-paraelectric material. Depth-resolved structures of the bilayers are measured with deep-sub-angstrom resolution and one picometer accuracy using multislice ptychography, enabling identification of the three-dimensional variations of polarization topology. Our findings reveal the evolution of the polar vortices in the twisted overlapping layers, demonstrating the reverse of rotation manner in the depth direction. Twisted freestanding bilayers provide a unique platform for exploration and modulation of novel polar topologies.**

Polar topologies play host to various novel physical properties, including enhanced conductivity[1,2], negative capacitance[3,4], sub-terahertz collective dynamics[5] and nonlinear light-matter interactions[6]. The design and detection of domains with distinct topological structures have become central topics in the field of ferroelectrics[7-9]. By now, various types of topological polar structures have been identified, such as flux-closure[10], vortex[11-13], skyrmion[14], meron[15], Solomon ring[16]. However, nearly all of these topological features are formed in ferroelectric thin films and/or grown on substrates. Polar topological structures in free-standing films are called for due to their increased flexibility[17] and additional degrees of freedom for controlling domain topology through stretching, bending, and torsion[8,9,18-20]. Additionally, the possibility of generating polar topologies in non-ferroelectric materials remains largely unexplored.

Simultaneously, moiré systems formed by 2D materials with a twisted angle have emerged as fertile grounds for exploring emergent novel phenomena[21-27]. Analogous to 2D atomic layers, freestanding oxide films with much flexibility[18,19,28,29] can also be employed to construct moiré patterns, which harbor the potential for inducing unique structures and properties. However, characterizing the interfaces and associated local structures within twisted bilayers poses a significant challenge. At present, atomic-scale polar displacements and nanoscale polarization and electric field can be determined by aberration-corrected scanning transmission electron microscopy (STEM)-based methods, like high-angle annular dark-field (HAADF) imaging and diffraction analysis based on 4D-STEM[3,4,14,17,30,31]. Nevertheless, most of these techniques can only acquire 2D projections of polarization structures, making it difficult to reveal the structure variation in the normal direction of twisted bilayers.

Ptychography is a high-resolution coherent diffractive imaging method used to retrieve object potentials encoded in diffraction intensities collected at series of probe positions (so-called 4D dataset)[32-35]. In multislice ptychography, samples are segmented into slices at various depths, and the potentials of each slice are reconstructed from the recorded coherent diffraction intensities[36,37]. Leveraging its advanced lateral resolutions and depth sectioning capabilities[38-42], multislice electron ptychography enables the exploration of three-dimensional structures concealed within twisted bilayers.

In this study, we demonstrate the occurrence of polar vortices in twisted freestanding bilayers of the paraelectric material, $SrTiO_3$, using multislice ptychography. The vortices are organized into in-plane arrays, while in the stacking direction, the vorticity alternates between clockwise and counterclockwise directions.

As shown in **Fig. 1**a, the twisted bilayer is formed by stacking two flakes of freestanding films of $SrTiO_3$. These films are grown on a $SrTiO_3$(001) substrate with a sacrificial layer of $Sr_2CaAl_2O_6$. The thickness of each freestanding layer is 17 nm. The low magnification annular dark field image (Extended Data Fig. S1) shows moiré fringes, which are characteristic of twisted bilayers. In order to get clear image of both stacking layers, HAADF is first used by tuning the beam defocus. As shown in Extended Data Fig. S2, the contrast decreases as the focal point of electron probe transitions from the upper layer to the lower layer. This contrast variation can be attributed to the channeling effect in the upper layer, yielding a much clearer contrast for the upper layer compared to the lower layer. Slight stretching of atomic columns in the HAADF image of the upper layer can be detected (the first image in Extended Data Fig. S2), which can be attributed to electron scattering of the lower layer. These observations indicate the challenge of decoupling the atomic structures of the two stacking layers, making accurate analysis of the bilayer's atomic structure a difficult task.

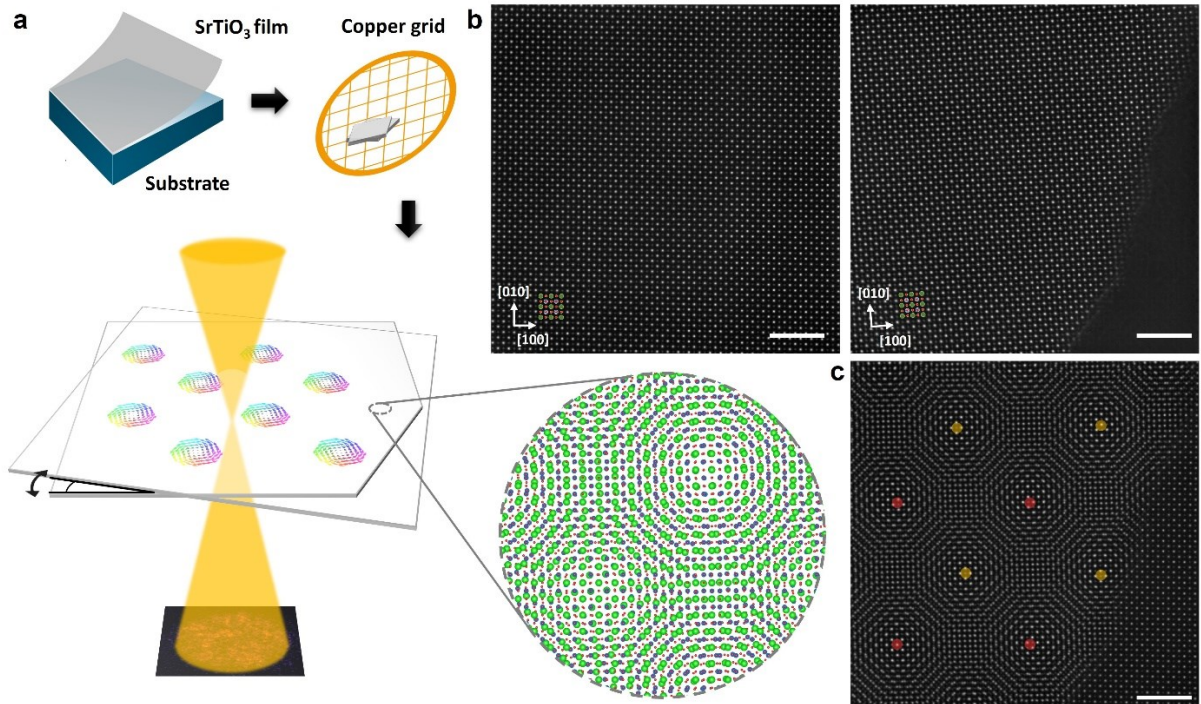

**Fig. 1. Ptychographic reconstruction of twisted SrTiO₃ bilayer. a,** Schematics illustrating the sample preparation process and experimental setup for electron ptychography. Besides is a structure model of twisted SrTiO₃. Sr, Ti, and O atoms are represented by green, blue and red spheres, respectively. **b,** Total phase images of the upper (left) and lower layer (right) recovered using multislice ptychography. Scale bar, 2 nm. Structure model of SrTiO₃ is overlaid on the phases. **c,** Total phase image summed over the slices of the two layers, revealing the moiré pattern in the stacking region. Regions with AA and AB stacking sequences are marked with yellow and red circles, respectively. Scale bar, 2 nm.

In contrast, multislice ptychography overcomes the channeling effect caused by dynamic scattering and enables the recovery of object functions for each depth section of the sample in the electron beam direction. During ptychographic reconstructions, the entire moiré system is divided into 4-Å-thick 'slices' for depth sectioning. Therefore, both upper and lower layers contain about 40 slices, ensuring identification of structural variations between and within the layers. Movie S1 provides the recovered phase images of all the slices. The integrated phase value of Sr and Ti columns are plotted as a function of depth (Extended Data Fig. S3) and used to evaluate the depth resolution, which is determined to be 2.2 nm (full width at 80% of maximum, FW80M). **Fig. 1**b presents the total phase of each layer, allowing for the measurement of a twist angle of 4.1°. The upper layer covers a larger region than the lower one, therefore the entire field of view encompasses both the twisted bilayer region and the single layer region. Moiré patterns are clearly visible in the total phase image obtained by summing the slices in the two layers (**Fig. 1**c). The coincidence regions in the moiré pattern are labelled by circles: those marked with yellow circles exhibit the AA (Sr-Sr, TiO-TiO) stacking sequence, maintaining the periodicity of bulk SrTiO₃, while regions marked with red circles display the AB (Sr-TiO, TiO-Sr) stacking sequence, resembling antiphase boundaries.

Based on the recovered phases at different depths, the positions of O and TiO columns in each slice are determined by Gaussian peak fitting. These positions were then used to calculate the polar displacements, which are defined as the relative shift between each TiO colum and its four nearest O colums ($\delta_{\text{O-Ti}}$). The polar displacements of all the slices in the upper layer can be seen in Extended Data Fig. S4, while those in the lower layer can be seen in Extended Data Fig. S5 Extended Data Movie S2 is used to show the variation process of the polarization in depth.

Polar vortices emerge in both upper and lower layers. In the upper layer, vortices are formed in depth of 11.2 nm and maintain such arrangement to the depth of 13.2 nm (Extended Data Fig. S4). The polarization mapping in **Fig. 2**a specifically focuses on the section at a depth of 12.0 nm. Here, the vortices exhibit counterclockwise (CCW) rotation. The vortices have a diameter of approximately 7~10 unit cells and are distributed between the regions with the same stacking sequence. **Fig. 2**b shows the vorticity, which is calculated as the z component of the curl of the polar displacement $(\nabla \times P)_{[001]}$. Vortices with CCW rotation form the regions with positive vorticity. Between the regions with different stacking sequences, vorticity is mainly negative. Moving to the lower layer, polar vortices form a regular array in the depth of 23.2 nm, approximately 7 nm below the interface (**Fig. 2**c and d). The core positions of the CCW vortices shift laterally compared to the upper layer, which can be caused by non-interface related strain field, such as nearby dislocations in the lower layer.

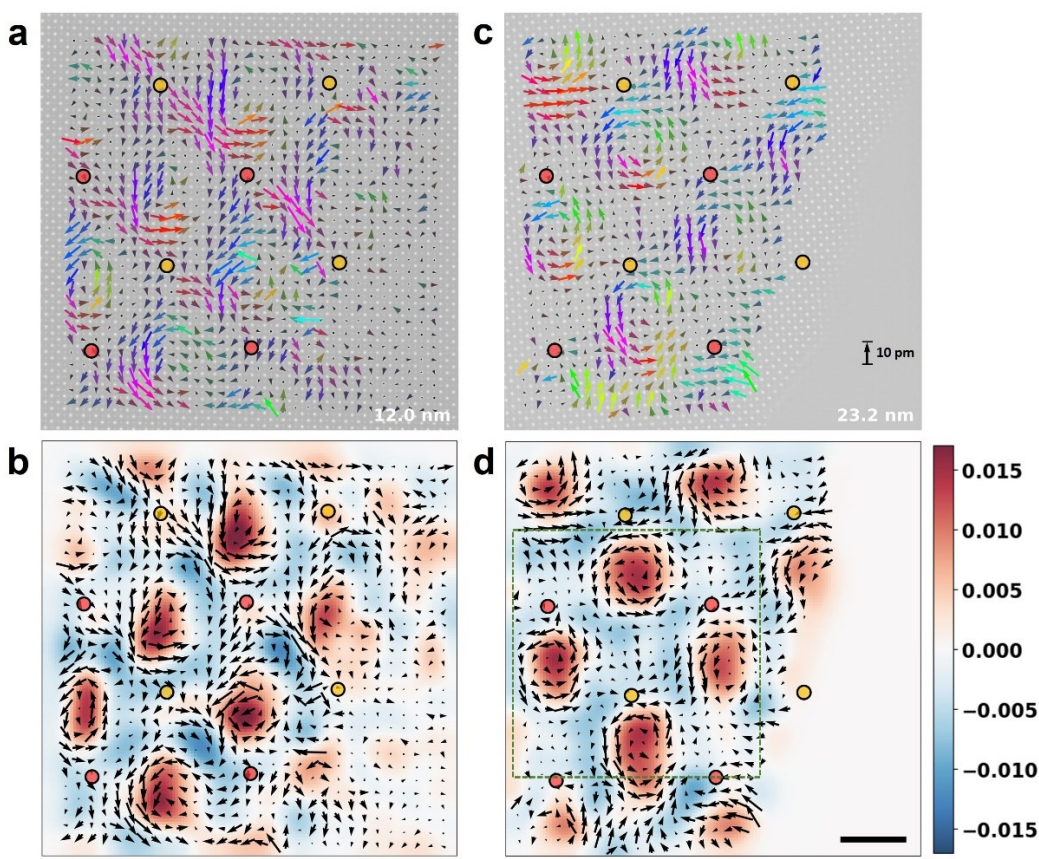

**Fig. 2. Polar vortices forming in the two stacking layers. a, c,** Polar displacement mapping in the depth of 12.0 nm (top layer) (a) and 23.2 nm (bottom layer) (c). Regions with AA and AB stacking sequences are marked with yellow and red circles, respectively. The 10-pm arrow in (c) is a reference for the magnitude of the polar displacement. **b, d,** Vorticity $(\nabla \times P)_{[001]}$ distribution corresponding to (a) and (c). The vorticity is in the unit of $10^{10}$ $C/m^3$. All the figures share the same scale bar, which is 2 nm shown in (d).

**Fig. 2**a reveals that polar vortices are only present in the bilayer region of the upper SrTiO$_3$ layer. The magnitude of the polar displacement is analyzed separately for the bilayer region and single-layer region of the upper layer. As Extended Data Fig. S6 shows, polar displacements in the single-layer region of the upper layer are much smaller compared to those in the twisted bilayer region. This observation suggests that the emergence of polar vortex is closely linked to the elastic interactions between the bilayers across the interface, which will be discussed in detail later. To provide a comparison, we conduct reconstruction on dataset simulated with rigid bilayers of $SrTiO_3$, which have no polarization and no elastic interactions across the interface. Simulation parameters such as the voltage, convergence semi-angle, scan step size, detector pixel size, and electron dose, are chosen to be the same as the experimental dataset. Extended Data Fig. S7 presents the mapping of polar displacement at two different depths near the interface. The polarization mapping of all the slices is shown in Movie S2. Neither the upper nor the lower layer exhibits polar topological features as seen in the experimental results. Extended Data Fig. S8a provides the magnitude of the relative displacements between TiO and O (or Sr) atomic columns ($\delta_{O-Ti}$ and $\delta_{Sr-Ti}$) calculated from all the recovered slices. The ground truth of $\delta_{O-Ti}$ and $\delta_{Sr-Ti}$ is zero, so the results can be viewed as the measurement accuracy. It can be observed that the values of $\delta_{O-Ti}$ and $\delta_{Sr-Ti}$ in the simulation are at the same level and smaller than the experimental ones, thereby confirming the existence of the polar vortex observed in the experiment. We also compared experimental values of $\delta_{Sr-Ti}$ and $\delta_{O-Ti}$. As shown in Extended Data Fig. S8b, $\delta_{O-Ti}$ is significantly larger than $\delta_{Sr-Ti}$, indicating the displacement of oxygen predominantly contributes to the formation of polar vortex.

For further validation of the reliability of ptychographic reconstruction, we performed experiments on a different bilayer sample with a distinct inter-layer rotation angle of 11.2° for comparison. Extended Data Fig. S9 shows the total phase images summed over the upper layer, lower layer and the entire sample. All the depth sections of polar displacements $\delta_{O-Ti}$, showcasing the absence of any polar vortex signature, are shown in Extended Data Movie S4.

Besides forming an in-plane vortex array, the polarization also evolves along the depth direction. **Fig. 3**a shows an evolution process of the reversal of vortex rotation ocurring in the lower layer. In the depth of 19.2 nm, polar vortices with CW rotation form a regular array (a local region is shown in the top left image of **Fig. 3**a and the whole region is shown in Extended Data Fig. S5). Some CCW vortices also exist near the CW vortices, creating pairs of vortices with opposite vorticity. As the depth increases, vortices first vanish and then reappear with the opposite rotation directions, forming a CCW vortex array in the depth of 23.6 nm (bottom right image of **Fig. 3**a and Extended Data Fig. S5). Throughout this process, the vortex cores remain nearly unchanged.

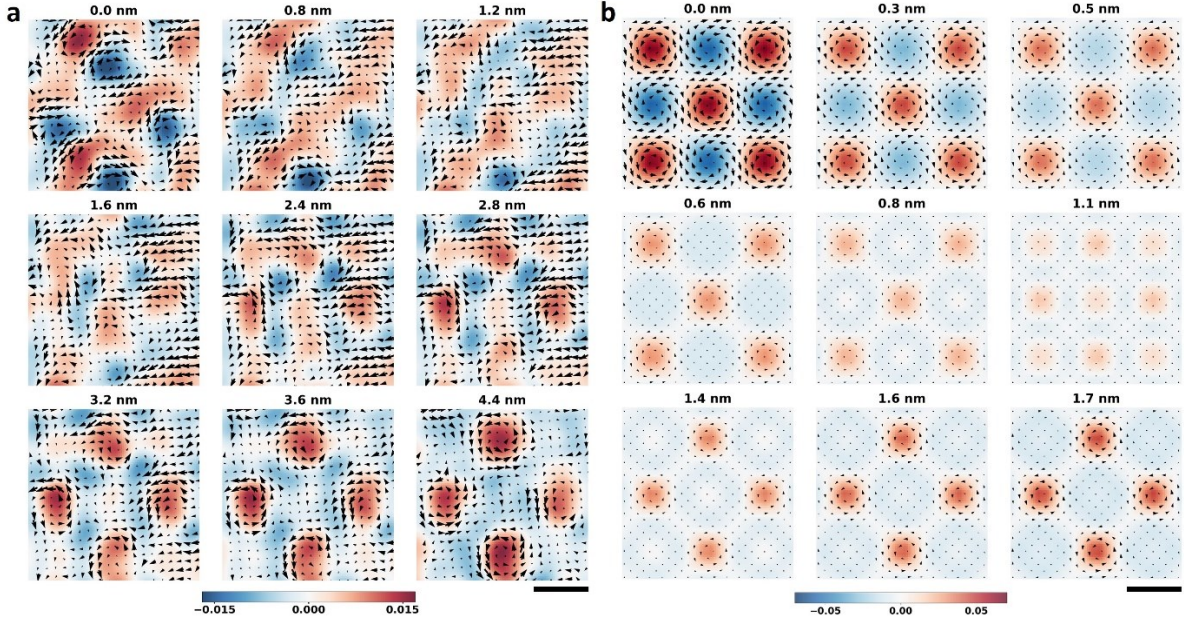

**Fig. 3. Evolving polar vortices in the depth direction. a,** Depth variation of the experimental polarization distribution in a small region in the lower layer. The corresponding regions in Fig. 2d are indicated by green squares. The relative depth values are labeled on the top of each image. Scale bar, 2 nm. **b,** Depth variation of the polarization from a phase-field simulation. Scale bar, 2 nm. The color maps of (a) and (b) stand for the vorticity in the unit of $10^{10}$ C/m$^3$.

To investigate the underlying mechanism of the polar vortex, we carried out phase-field modeling (details in Methods). Our findings reveal that incorporating flexoelectricity is necessary to reproduce the experimental observations, indicating its significant role in introducing polar vortex in paraelectric SrTiO$_3$. Extended Data Fig. S10 shows the detailed influence of flexoelectric coefficients on the polarization distribution in phase-field simulation. Previous studies have demonstrated the importance of flexoelectricity in introducing polarization in defects[43,44] and superlattice[45]. The flexoelectric coefficients used in our simulation ($f_{11}$ = 1 V, $f_{12}$ = 0 V, $f_{44}$ = 3 V) align well with the experimental measurements[46,47]. **Fig. 3**b shows the depth sections of polarization in a region with identical size as presented in **Fig. 3**a, which bears a close resemblance to the experimental observation.

The distribution of polarization is found to be closely tied to the lattice rotation. **Fig. 4**a shows the distribution of lattice rotation (measured using geometric phase analysis[48]) and the corresponding polarization displacement in the depth of 19.2 and 23.2 nm in the lower layer. Most CCW vortices are located in the region with CCW lattice rotation, while CW vortices are located in the region with CW lattice rotation. The relationship is futher highlighted through the line profiles of vorticity and rotation in **Fig. 4**b. As the direction of the vortices reverses, so does the lattice rotation. This regularity is validated by phase-field modeling. **Fig. 4**c shows the simulated lattice rotation and polar displacement at two different depths, while **Fig. 4**d presents their line profiles. The concurrent change in the sign of vorticity and lattice rotation matches well with the experiment, confirming the strong correlation between the directions of vortex and lattice rotations.

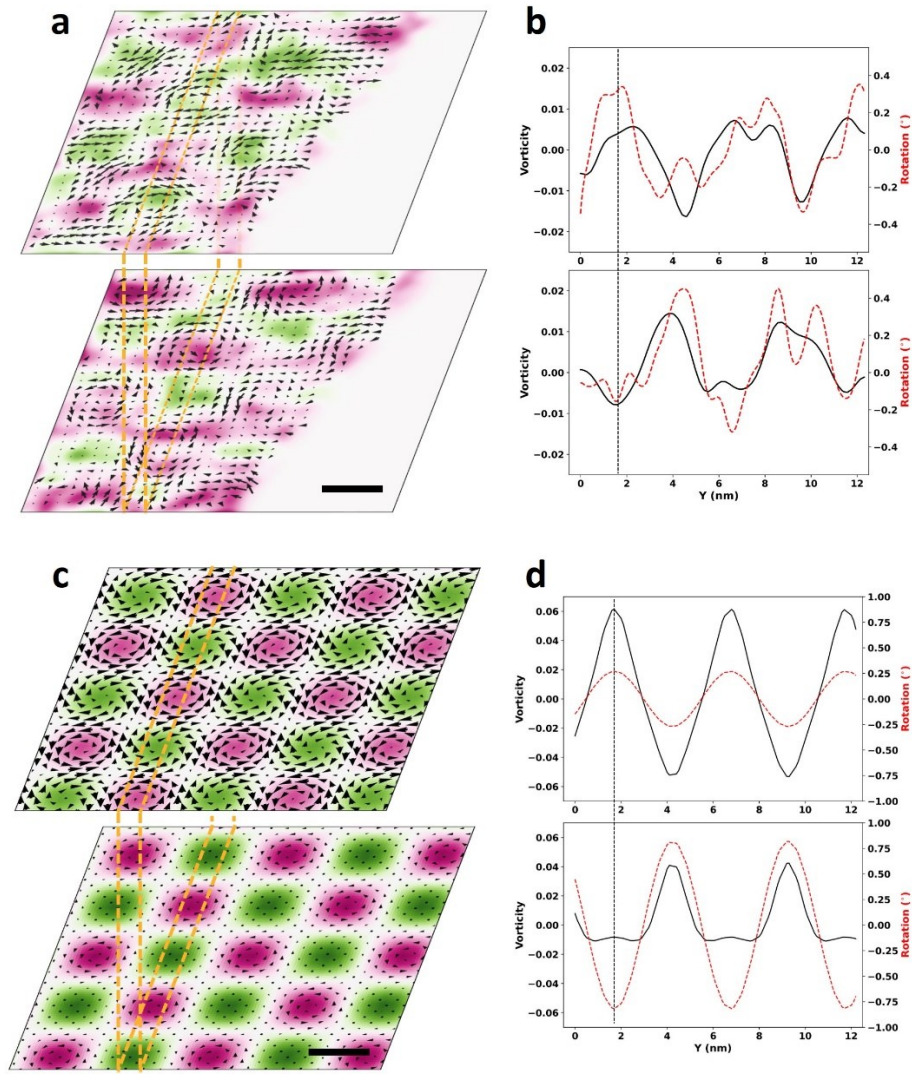

**Fig. 4. Relationship between polar vortex and lattice rotation. a, c,** Phase-field modeling (a) and experimental results (c) of lattice rotation and polarization in two depths with CW and CCW vortex arrays. **b**, **d,** Profiles of the vorticity and lattice rotation across the lines marked in (a) and (b). Positive values mean counterclockwise lattice rotation. Scale bars, 2 nm.

This study unravels the formation of polar vortices in the twisted film of paraelectric $SrTiO_3$. Using multislice electron ptychography, the structures of the overlapping layers are distinguished, allowing the study of the 3D variation of polarization in each layer. It is found that polar vortices evolve into different rotation manners in the depth direction, and the emergence of vortex array is tightly related to the lattice rotation. Our work demonstrates that with the existence of flexoelectric effect, polar topologies can also be generated in paraelectric materials. Findings in this work indicates some unique advatanges for potential applications of twisted freestanding $SrTiO_3$ in non-volatile memories and microwave devices (**Fig. 5**). The small size of the vortices (~4 nm) is promising to realize high-density memories (**Fig. 5**a). Besides the feasibility of freestanding film, twist angle will act as a new freedom to increase the tunability of the system. Theoretically, properties such as vorticity, spacing between vortices and vortex size are adjustable, which is significant to tailor device performance (**Fig. 5**b). Also, lower energy barrier between different states of paraelectric SrTiO$_3$ will lead to low power usage (**Fig. 5**c). At last, it will be easy to integrate the twisted freestanding film to silicon-based device.

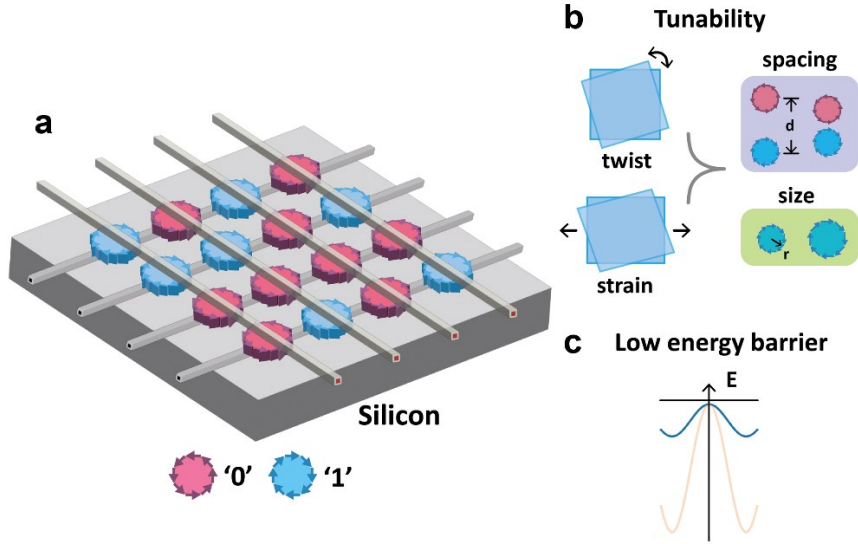

**Fig. 5. Potential application and advantages of twisted freestanding SrTiO$_3$. a,** Memory devices with high storage density. **b,** High tunability. Twisting and strain are easily applied to the system, which can be used to manipulate the vortex geometries. **c,** Low energy barrier between different states compared to conventional ferroelectric materials.

**Methods**

Preparation of twisted bilayer SrTiO$_3$ film

Pulsed laser deposition (248 nm KrF excimer laser) was used to grow the epitaxial heterojunctions SrTiO$_3$/Sr$_2$CaAl$_2$O$_6$ on SrTiO$_3$ (001) substrates. The substrate temperature and oxygen partial pressure used to grow the Sr$_2$CaAl$_2$O$_6$ (SAO) layer is 740 °C and 15 Pa, respectively. The substrate temperature and oxygen partial pressure used to grow the SrTiO$_3$ film was 710 °C and 10 Pa, respectively. During the deposition, the energy density of laser beam focused on the targets was 2 J cm$^{-2}$ and 1.9 J cm$^{-2}$ for Sr$_2$CaAl$_2$O$_6$ and SrTiO$_3$, respectively. The laser pulse rate was 3 Hz. The sample was annealed at 710 °C for 10 min under an oxygen partial pressure of 100 Pa.

To transfer the freestanding film to TEM grid, polymethyl-methacrylate (PMMA) was coated onto the as-grown film. Tape was then attached to the surface of PMMA. Then, the sample was immersed in deionized water at 120 °C until the SAO layer was dissolved. During this process, the whole film cracked into pieces and some of them attached to the surface of others. Next, the surface of the film was attached to the carbon film side of a TEM grid and the PMMA outside of the grid was cut. At last, the PMMA was removed by acetone vapor.

Experimental data collection and reconstruction

The electron microscopy experiment was conducted in a probe aberration-corrected FEI Titan Cubed Themis G2 operated at 300 kV. The convergence semi-angle is 25 mrad. The nominal collection angle range of the ADF image shown in Fig. 1B is 25~153 mrad. 4D datasets used to do multislice ptychography are acquired using an electron microscope pixel array detector (EMPAD). Each diffraction pattern contains 128×128 pixels and the pixel size is 0.026 Å$^{-1}$. The probe is under-focused about 30 nm and the scan step size is 0.52 Å. The dataset was acquired under a beam current of about 20 pA and a dwell time of 0.1 ms, which is measured by EMPAD.

A homemade program named EMPTY (Electron Microscopic PTYchography) mentioned in our previous work[40] was used for the ptychographic reconstruction. When doing multislice ptychography, diffraction patterns were padded with zero to 160×160 pixels to generate a real space pixel size (sampling interval of recovered object phases) of 0.24 Å. Object was separated into 85 slices and the slice thickness was 4 Å. Six probe states were used to account for the partial coherence of the illumination[49,50]. Probe positions were refined using the gradient descent method[51]. The regulation factor used to avoid ambiguity in multislice electron ptychography is initialized as 1.0 and gradually decreased after reaching convergence.

Simulation of ptychography dataset

A homemade multislice simulation program was used to generate the 4D dataset used for ptychographic reconstruction. The reliability of the program was verified in our previous work. The voltage, convergence angle, and pixel size of diffraction patterns are the same as in the experiment. The thickness of the twisted SrTiO$_3$ bilayer model is 30 nm in total, which is close to the sample used in the experiment. The scan step size is 0.468 Å. Poisson noise was added to diffraction patterns, corresponding to a beam of 30 pA and a dwell time of 1 ms. During reconstruction, the slice thickness of the sample is also 4 Å.

Analysis of reconstructed phase images

The Python package Atomap was used to locate atomic columns in phase images. After finding Sr, TiO, and O columns, the polar displacement of each unit cell was calculated as the deviation between the TiO column and the center of four neighboring O columns. Geometric phase analysis was calculated using Strain++.

Phase-field simulation

A three-dimensional ferroelectric polarization vector $\boldsymbol{P} = (P_x, P_y, P_z)$ was chosen as the order parameter. The temporal evolution of $\boldsymbol{P}$ was obtained by solving the time-dependent Ginzburg-Landau equation:

$$\frac{\partial P_i(\boldsymbol{r},t)}{\partial t} = -L \frac{\delta F}{\delta P_i(\boldsymbol{r},t)} (i = x, y, z)$$

in which $L$ denotes the kinetic coefficient, $F$ denotes total free energy functional, $\boldsymbol{r}$ denotes spatial coordinate, and $t$ denotes evolutionary time. Meanwhile, total free energy functional $F$ includes contributions of Landau energy, gradient energy, electric energy, flexoelectric energy, elastic energy, and electrostrictive energy respectively:

$$F = \iiint \left( f_{Landau} + f_{Gradient} + f_{Electric} + f_{Flexo} + f_{Elastic} + f_{Es} \right) dV$$

which can be given by

$$f_{Landau} = \alpha_{ij} P_i P_j + \alpha_{ijkl} P_i P_j P_k P_l + \alpha_{ijklmn} P_i P_j P_k P_l P_m P_n + \alpha_{ijklmnrs} P_i P_j P_k P_l P_m P_n P_r P_s$$

$$f_{Gradient} = \frac{1}{2} G_{ijkl} P_{i,j} P_{k,l}$$

$$f_{Electric} = -\frac{1}{2} \kappa_0 E_i E_i - P_i E_i$$

$$f_{Flexo} = -\frac{1}{2} f_{ijkl} (\varepsilon_{ij,l} P_k - \varepsilon_{ij} P_{k,l})$$

$$f_{Elastic} = \frac{1}{2} C_{ijkl} \varepsilon_{ij} \varepsilon_{kl}$$

$$f_{Es} = -q_{ijkl} \varepsilon_{ij} P_k P_l$$

where $\alpha_{ij}$, $\alpha_{ijkl}$, $\alpha_{ijklmn}$ are Landau coefficients, $G_{ijkl}$ gradient coefficients, $f_{ijkl}$ flexoelectric coefficient, $C_{ijkl}$ electrostrictive coefficient, $q_{ijkl}$ elastic constant, and $\kappa_0$ background dielectric permittivity. $\varepsilon_{ij}$ denotes the strain component and $E_i$ the electric field component derived from $E_i = \varphi_{,i}$, where $\varphi$ is the electric potential. Each energy expression can be found in previous literature. In addition, the mechanical ($\sigma_{ij,j} = 0$) and electric ($D_{i,i} = 0$) equilibrium conditions must be satisfied for the body-force-free and body-charge-free systems in simulation. A finite element method was employed to solve the above equations to obtain spatio-temporal evolution of polarization, stress, and electric field.

For simplicity, a set of $500\Delta x \times 500\Delta x \times 20\Delta x$ uniform meshing is used in this model, where $\Delta x$ represents 0.5 nm. Periodic boundary conditions are set up along the in-plane directions ([100] and [010]). A close-circuit electric boundary condition was applied on both top and bottom surfaces. A fixed displacement boundary condition was employed on the bottom of the film, while the stress on the top surface was free. The amplitude of the initial random noise of $P$ was set as 0.001 C/m$^{-2}$ for the following polarization evolution.

To simulate the periodic strain field in twisted bilayers, the estimated in-plain strain was applied with sinusoidal waveform, which can be derived from elastic theory:

$$\varepsilon_{xx} = \varepsilon_{yy} = -\frac{A}{2} \cdot \left[ \cos\left(\frac{2\sqrt{2}\pi}{k_x l_0} x\right) + \cos\left(\frac{2\sqrt{2}\pi}{k_y l_0} y\right) \right]$$

$$\varepsilon_{xy} = \frac{A}{2} \cdot \left[ \cos\left(\frac{2\sqrt{2}\pi}{k_x l_0} x\right) - \cos\left(\frac{2\sqrt{2}\pi}{k_y l_0} y\right) \right]$$

where $A$ is the strain amplitude, $k_x l_0$, and $k_y l_0$ are periods of the in-plain strain of which $l_0$ is the lattice parameter. According to the strain in experiments, $A = 0.015$ and $k_x = k_y = $ were assumed to achieve the same vorticity and period of vortices.

**Acknowledgments:** In this work we used the resources of the Physical Sciences Center and Center of High-Performance Computing, Tsinghua University.

**Funding:** This work was supported by the National Natural Science Foundation of China (52388201, 51525102).

**Author contributions:** R.Y. and H.H. designed and supervised the research. H.S. performed ptychography experiment, reconstruction and analysis. Y.Z. performed phase field simulations. H.S. and J.C. performed diffraction simulations. W.Y. assisted with experiments and data analysis. Y.M. and Q.L. grew the free-standing SrTiO$_3$ film. H.S., R.Y., Y.Z., and H.H. co-wrote the paper. All authors discussed the results and commented on the manuscript.

**Competing interests:** Authors declare that they have no competing interests.

**Data and materials availability:** All data are available in the main text or the supplementary materials.

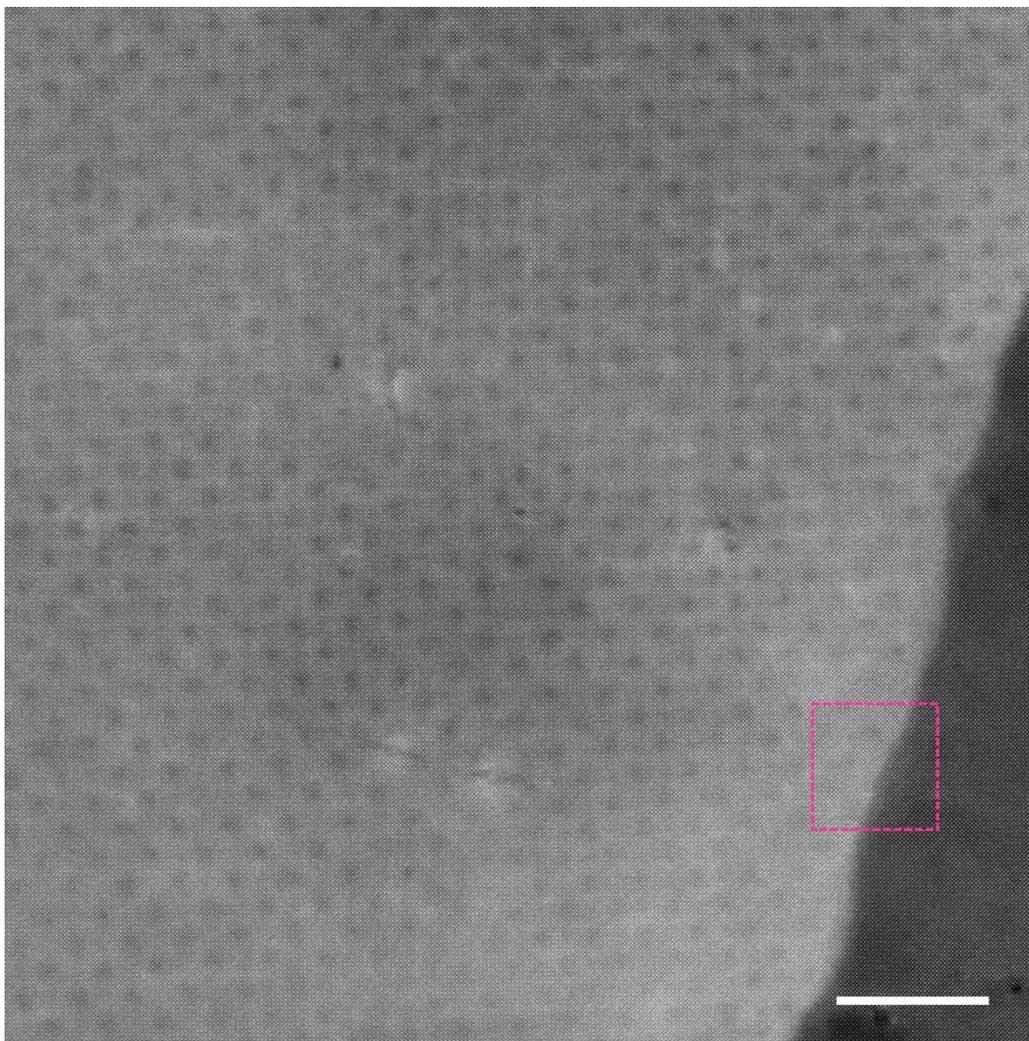

Extended Data Fig. S1. Low magnification ADF image. Scale bar, 15 nm. The region marked by red dashed box is reconstrued by ptychography, which is shown in Fig. 1(b) and (c).

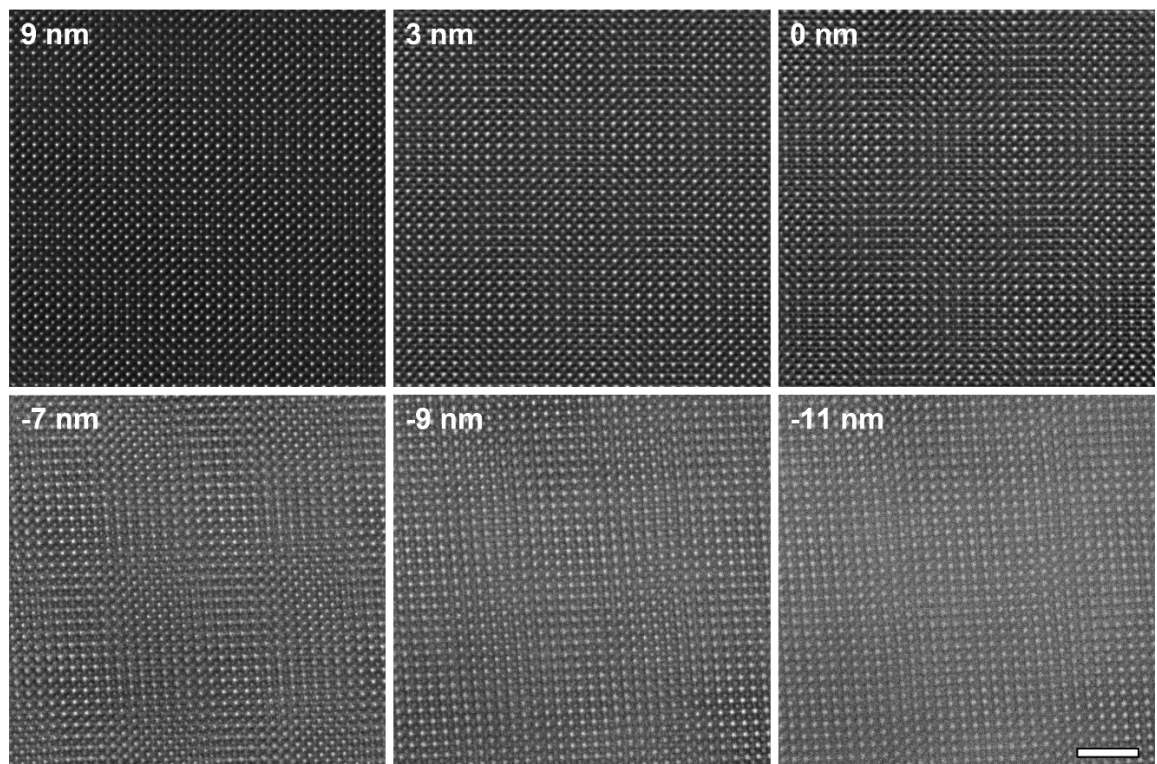

Extended Data Fig. S2. A focal series of Wiener-filtered HAADF images. The relative values of beam defocus are labeled on each image. A positive value stands for overfocus. All the images are acquired by multi-frame fusion (40 frames in total) with drift correction offered by Velox software. Dwell time is 150 ns (0.7 s per frame). Scale bar, 2 nm.

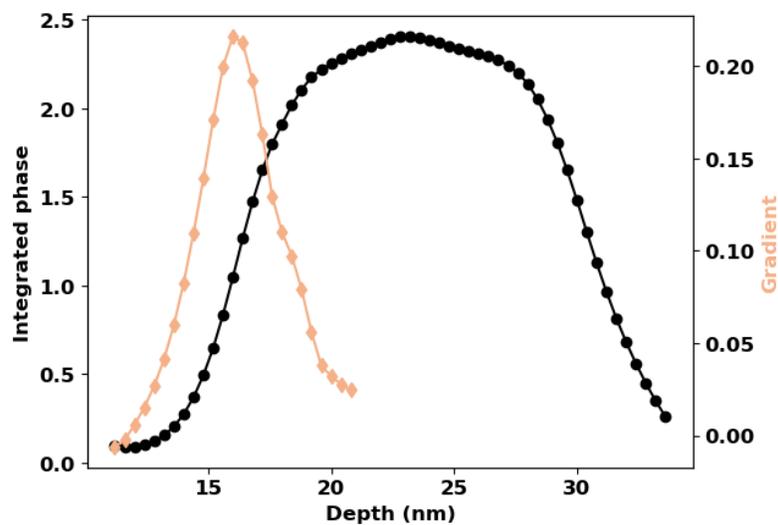

Extended Data Fig. S3. Depth profile of the integrated phase of Sr and Ti columns (black circle). The gradient of the depth profile is shown with pink diamond. Gaussian peak is used to fit the gradient curve and FW80M is calculated as the depth resolution.

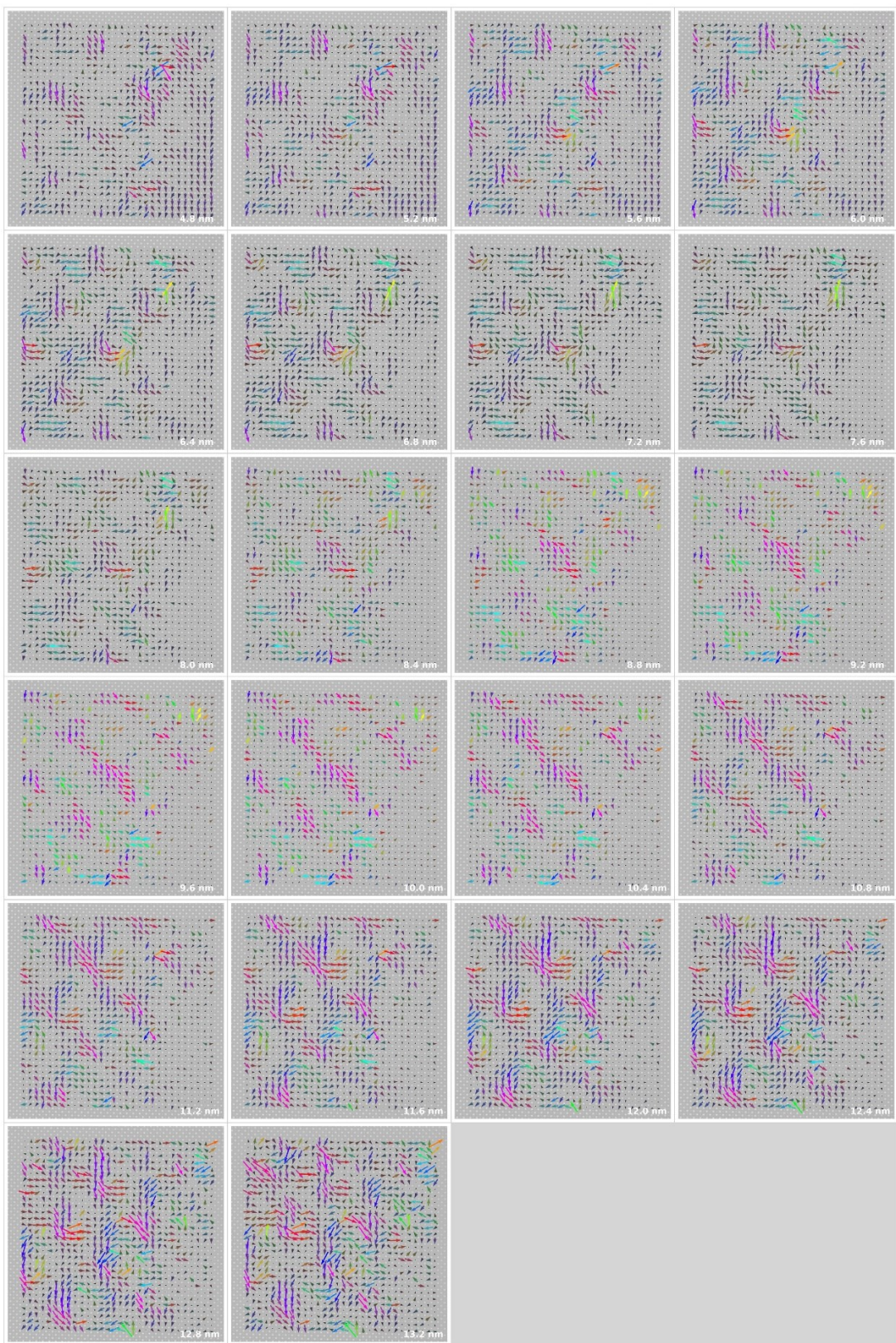

Extended Data Fig. S4. Distribution of in-plane polarization at different depths in the upper $SrTiO_3$ layer. Distances to the upper surface are labeled in the lower left corner.

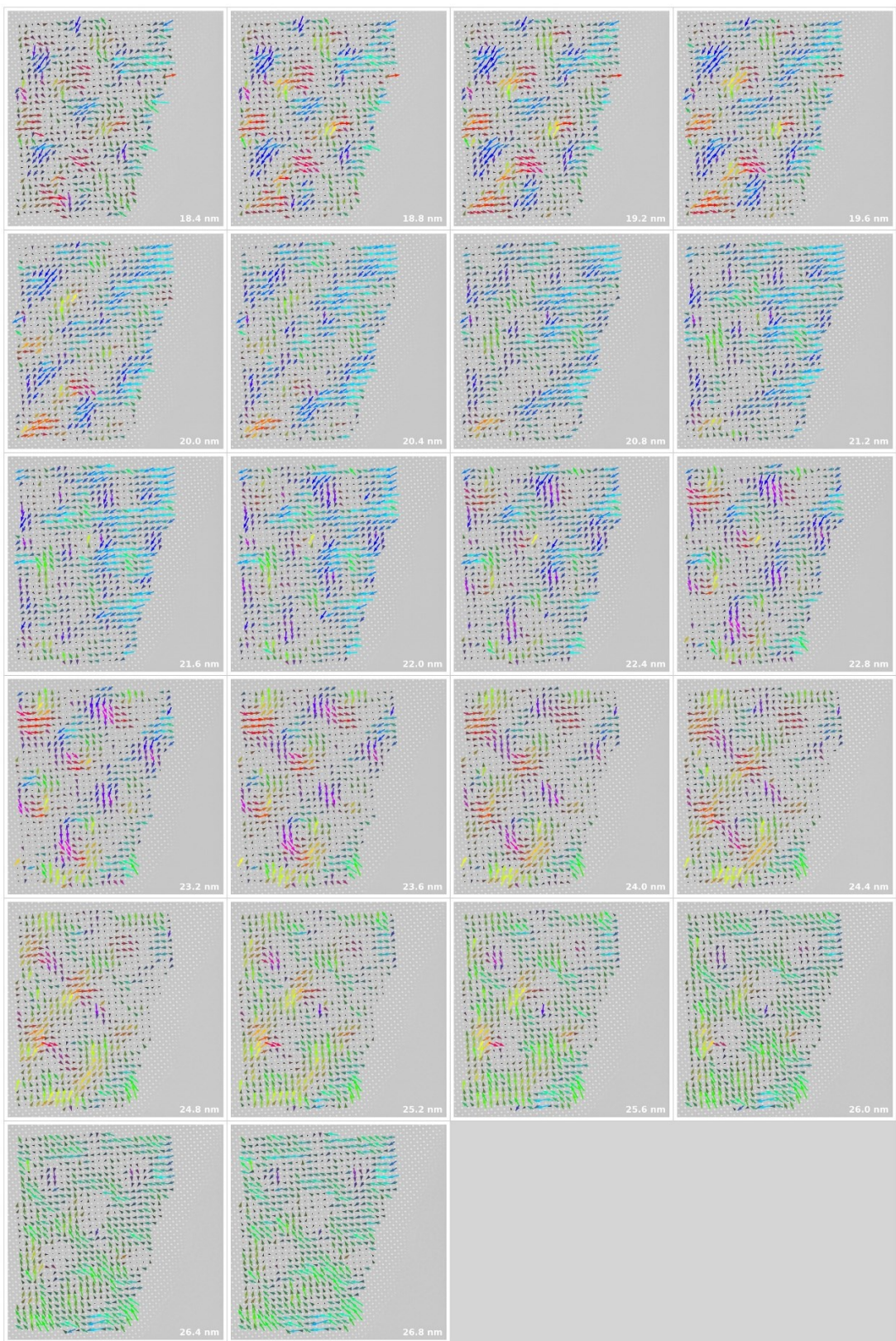

Extended Data Fig. S5. Distribution of in-plane polarization at different depths in the lower $SrTiO_3$ layer. Distances to the upper surface of the upper layer are labeled in the lower left corner.

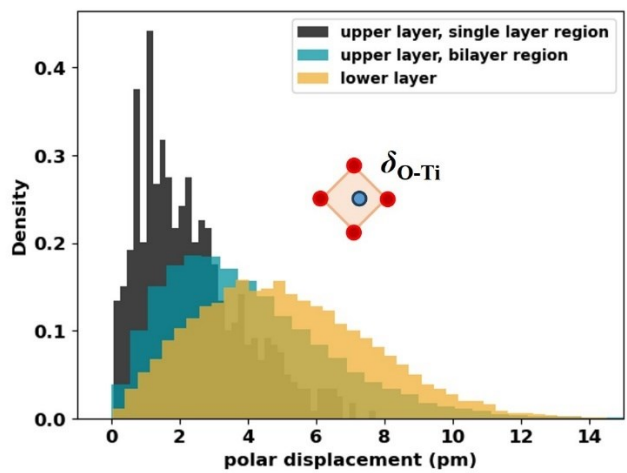

Extended Data Fig. S6. Histogram of the relative displacement ($\delta_{\text{O-Ti}}$) in experiment.

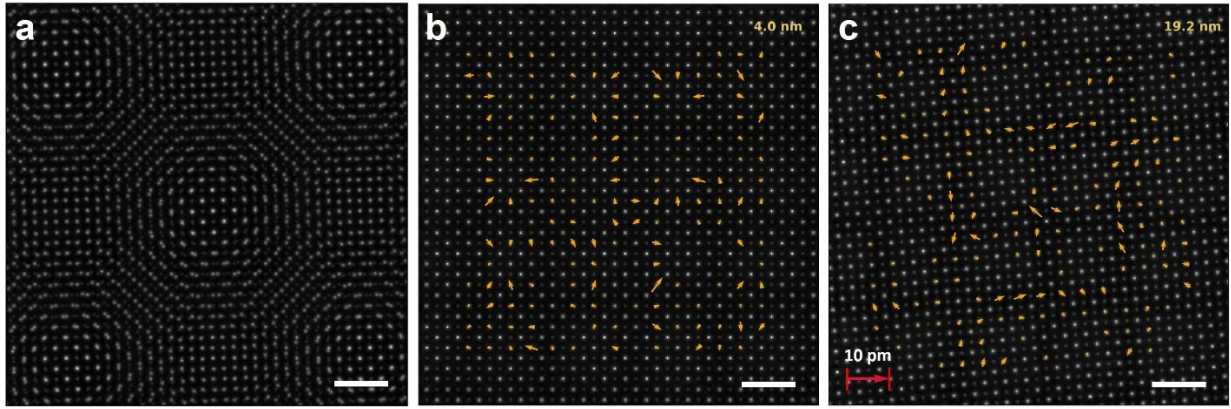

Extended Data Fig. S7. Phases of twisted SrTiO$_3$ reconstructed using simulated datasets. (a) Total phase image summed over all the recovered slices. (b) Polar displacement in the depth of 4 nm inside the upper layer. (c) Polar displacement in the depth of 19.2 nm inside the lower layer. The red arrow stands for 10-pm polar displacement. 4D datasets are simulated using the same parameters as experiment. Scale bars, 1 nm.

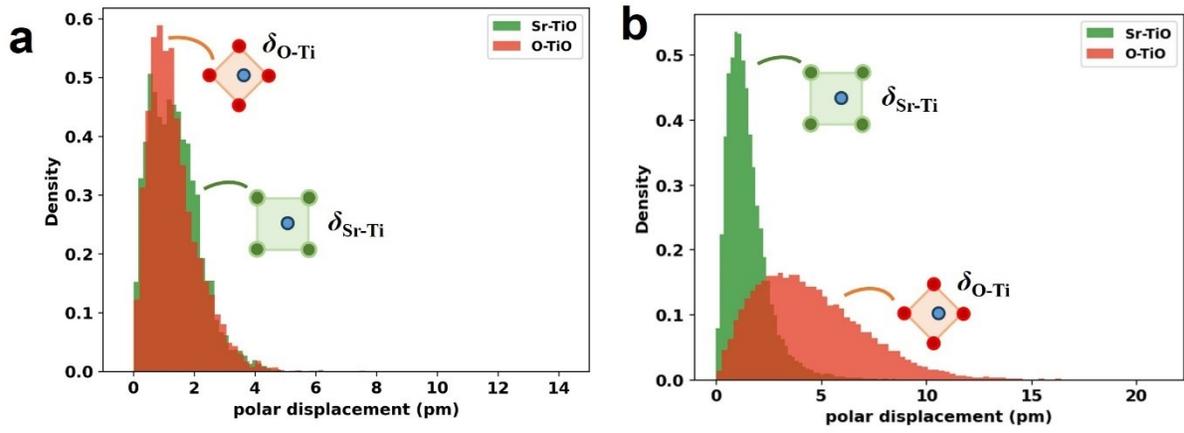

Extended Data Fig. S8. Comparison of relative displacements calculated with different methods ($\delta_{\text{O-Ti}}$ and $\delta_{\text{Sr-Ti}}$). Results in experiment (a) and simulation (b) are shown separately.

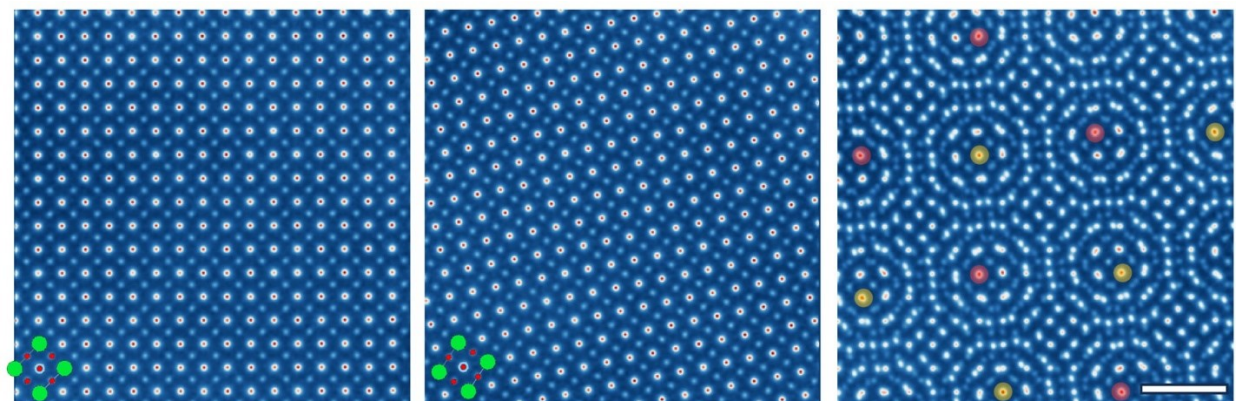

Extended Data Fig. S9. Experimental total phases of the upper layer (left), lower layer (middle), and the double layers (right) reconstructed from the sample with an inter-layer rotation angle of 11.2°. Scale bar, 1 nm. $SrTiO_3$ structure model is overlaid on the phase images. The yellow and red circles mark the region with AA (Sr-Sr, TiO-TiO) and AB (Sr-TiO, TiO-Sr) stacking sequence, respectively.

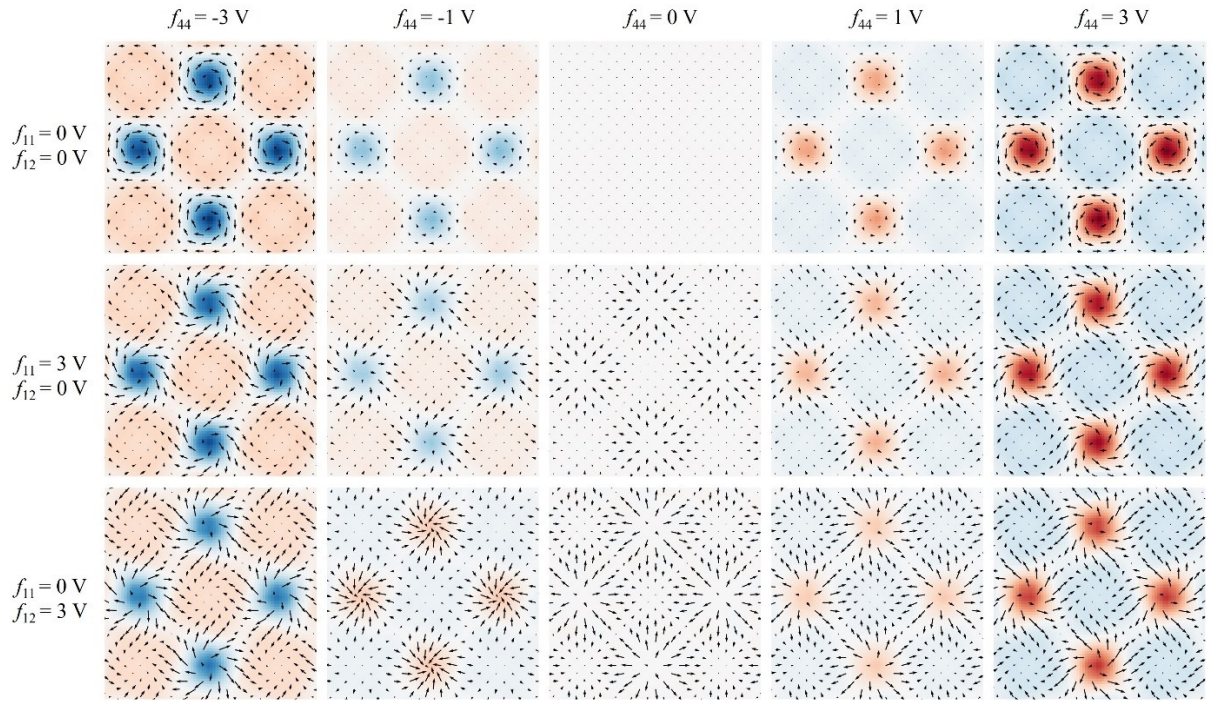

Extended Data Fig. S10. The evolution of the vortex patterns in dependent with longitudinal ($f_{11}$), transverse ($f_{12}$), and shear ($f_{44}$) flexoelectric effects. It can be seen that no in-plain vortex appears when all flexoelectric coefficients are zero(middle image in the first line). Shear flexoelectric coupling(with coefficient $f_{44}$) mainly resulted in the vortices, while longitudinal ($f_{11}$) and transverse ($f_{12}$) ones produced uneven in-plain polar textures, which distorted formed vortices.

Extended Data Movie S1. Phase images of all the recovered slices.

Extended Data Movie S2. Polar displacement mappings of the upper and lower layers of the sample with 4.1° inter-layer rotation in experiment.

Extended Data Movie S3. Polar displacement mappings of the upper and lower layers in simulation.

Extended Data Movie S4. Polar displacement mappings of the upper and lower layers of the sample with 11.2° inter-layer rotation angle in experiment.